\documentclass[preprint,preprintnumbers]{revtex4-1}
\usepackage{graphicx}
\usepackage{dcolumn}
\usepackage{bm}
\usepackage{amsmath}
\usepackage{amsfonts}
\usepackage{rotating}
\usepackage{xspace}
\usepackage{color}
\usepackage{float}

\newcommand{\cm}{\ensuremath{\rm cm^{-1}}\xspace}

\begin{document}

\title{Lifetimes and wave functions of ozone metastable vibrational states  near the dissociation limit in full symmetry approach}
\author{David Lapierre$^{1}$, Alexander Alijah$^{1}$, Roman Kochanov$^{2,3}$,  Viatcheslav Kokoouline$^{4}$, and  Vladimir Tyuterev$^{1}$}
\affiliation{$^{1}$Groupe de Spectrometrie Mol\'eculaire et Atmospherique, UMR CNRS 7331, 
University of Reims Champagne-Ardenne, F-51687, Reims Cedex 2, France \\
$^{2}$Quamer Laboratory, Tomsk State University, Tomsk, Russia \\
$^{3}$Harvard-Smithsonian Center for Astrophysics
Atomic and Molecular Physics Division, MS50 60 Garden St, Cambridge MA 02138  USA\\
$^{4}$Department of Physics, University of Central Florida, Orlando, FL 32816, USA
 }


\begin{abstract} 
 Energies and lifetimes (widths) of vibrational states above the lowest dissociation limit of $^{16}$O$_3$ were determined using a previously-developed efficient approach, which combines hyperspherical coordinates and a complex absorbing potential. The calculations are based on a recently-computed potential energy surface of ozone determined with a spectroscopic accuracy [J. Chem. Phys.  {\bf 139}, 134307 (2013)]. The effect of permutational symmetry on rovibrational dynamics and the density of resonance states in O$_3$  is discussed in detail. Correspondence between quantum numbers appropriate for  short- and long-range parts of wave functions of the rovibrational continuum is established. It is shown, by symmetry arguments, that the allowed purely vibrational ($J=0$) levels of $^{16}$O$_3$ and  $^{18}$O$_3$, both made of bosons with zero nuclear spin, cannot dissociate on the ground state potential energy surface. Energies and wave functions of bound states of the ozone isotopologue $^{16}$O$_3$ with rotational angular momentum $J=0$ and 1 up to the dissociation threshold were also computed. For bound levels, good agreement with experimental energies is found: The RMS deviation between observed and calculated vibrational energies  is 1~\cm. Rotational constants were determined and used for a simple identification of vibrational modes of calculated levels.
\end{abstract}


\maketitle


\section{Introduction}

Knowledge of quantum rovibrational states near the dissociation threshold is mandatory for the understanding 
of the molecular dynamics of formation and depletion processes.  In this respect the ozone molecule is a 
particular interesting subject for both fundamental molecular 
physics~\cite{SCH06:625, MAR13:17703,IVA13:17708,SUN10:555,LI14:081102,RAO15:633,GRE07:2044,TYU14:143002,GAR06:163005,MAU16:054107}
and various applications owing to the well-known role that this molecule plays in atmospheric physics and 
climate processes~\cite{LU09:118501,BOY09:6255}. 
Despite of the significant progress made over past decades in the study of ozone 
spectroscopy~\cite{TYU14:143002,flaud1990atlas,MIK96:227,CAM06:1,BAR13:172,MON12:840,BAB14:169} 
and dynamics~\cite{SCH06:625, MAR13:17703,IVA13:17708,SUN10:555,LI14:081102,RAO15:633,MAU16:054107,XIA89:6086,GAO01:259,
GAO02:137,CHA04:2700,XIE05:131,WIE97:745,SCH03:1966,LIN06:5305,VAN07:2866,VET07:138301,GRE09:181103,DAW11:081102}
many aspects of this molecule as well as of the $\rm O_2 + O$ complex in high energy states are not yet fully understood.  
One of the major motivations for recent  investigations of excited ozone has been  the discovery of the mass-independent 
fractionation reported by  Mauersberger {\it et al.}~\cite{MAU81:935,KRA96:1324,JAN03:34}, 
Thiemens {\it et al.}~\cite{THI83:1073}, Hippler {\it et al.}~\cite{HIP90:6560},
in laboratory and atmospheric experiments: for most molecules, the isotope enrichment scales 
according to relative mass differences, but the case of ozone shows an extremely marked deviation from this rule. 
This has been considered as a ``milestone in the study of isotope effects''~\cite{MAR13:17703}  and 
a ``fascinating and surprising aspect $\dots$ of selective enrichment of heavy ozone isotopomers''~\cite{DAW11:081102}.
On the theoretical side, many efforts have been devoted to the interpretation of these findings, 
in the research groups of Gao and Marcus~\cite{GAO01:259,GAO02:137}, Troe {\it et al.}~\cite{LUT05:2764,HIP90:6560}, 
Grebenshchikov and Schinke~\cite{GRE09:241,GRE09:181103}, 
Babikov {\it et al.}~\cite{VET07:138301,BAB03:2577},  Dawes {\it et al.}~\cite{DAW11:081102} 
and in many other studies, see~\cite{SCH06:625, MAR13:17703,IVA13:17708,SUN10:555,LI14:081102,RAO15:633,CHA04:2700,XIE05:131,WIE97:745,SCH03:1966,LIN06:5305,VAN07:2866,JAN01:951,GUI15:12512,NDE15:12043,NDE15:7712}
and references therein.  
Several fundamental issues raised by the ozone studies could have an impact on the understanding 
of important phenomena in quantum molecular physics and of the complex energy transfer dynamics near the dissociation threshold. 

It has been recognized that a non-trivial account of the symmetry properties~\cite{RAO15:633,JAN01:951}, 
efficient variational methods for the nuclear motion calculations and an accurate determination of the 
full-dimensional ozone potential energy surfaces (PES) are prerequisites for an adequate description 
of related quantum states and processes in the high energy range. The ozone molecule exhibits a complex 
electronic structure and represents a challenge for accurate ab initio 
calculations~\cite{GRE07:2044,GAR06:163005,XAN91:8054,SIE01:1795,SIE02:9749,KAL08:054312,HOL10:9927,SHI11:184104}.
Earlier  1D PES studies predicted an ``activation barrier'' at the transition state (TS) along the 
minimum energy path (MEP)~\cite{ATC97:176,HAY82:862,BAN93:155}. 
Later on more advanced electronic structure calculations have 
suggested that the MEP shape could have a ``reef''-like structure~\cite{HER02:478,SCH04:5789,FLE03:610} 
with a submerged barrier below the dissociation limit. Following preliminary estimations of 
Fleurat-Lessard {\it et al.}~\cite{FLE03:610}, this ``reef'' 
feature was incorporated into a so-called ``hybrid PES'' by Babikov {\it et al.}~\cite{BAB03:2577} by introducing 
a 1D semi-empirical correction to the three-dimensional Siebert-Schinke-Bittererova (SSB)~\cite{SIE01:1795,SIE02:9749}  PES 
with empirical adjustments to match the experimental dissociation energy. 
This modified mSSB surface containing a shallow van der Waals (vdW) minimum along the dissociation 
reaction coordinate around $r_1 \sim 4.5 - 5.0 \, a_0$  has been used to study the metastable 
states~\cite{BAB03:2577} and also suggested the existence of van der Waals 
bound states~\cite{GRE03:6512,JOY05:267,BAB03:6554,LEE04:5859}. 
A detailed review of ozone investigations up to  this stage has been presented in the 
``Status report of the dynamical studies of the ozone isotope effect'' by Schinke {\it et al.}~\cite{SCH06:625} 
who concluded that the calculated rate constants were about 3-5 times smaller than the measured ones and had 
a wrong temperature dependence. 
Recently Dawes {\it et al.}~\cite{DAW11:081102,DAW13:201103} have argued that an accurate account of several interacting 
electronic states in the TS region should result in a ground state potential function without the 
``reef'' feature found in previous ab initio calculations.  Since this work, 
and based on scattering studies~\cite{LI14:081102,XIE15:064308,SUN15:174312}, 
the ``reef structure'' was considered a ``deficiency''~\cite{NDE16:074302} of the SSB PES~\cite{SIE01:1795,SIE02:9749} 
and its modified mSSB versions~\cite{BAB03:2577,AYO13:164311}, 
and was thought a plausible reason for the disagreement in rate constant calculations~\cite{SCH06:625,DAW11:081102}. 
Ndengue {\it et al.}~\cite{NDE16:074302} have reported energies of  $J = 0$ and $J = 1$ bound rovibrational ozone states 
below $D_0$ using the Dawes {\it et al.} \cite{DAW13:201103} PES. Variational calculations of the 100 lowest bound vibrational states
using that PES resulted in a root-mean-square (RMS) obs-calc error~\cite{NDE16:074302} 
of  $\sim 20 \, \cm$  with respect to the experimentally observed band centers of $\rm (^{16}O)_3$.  

In 2013 Tyuterev {\it et al.}~\cite{TYU13:134307} have proposed a new analytical representation for the ozone PES accounting 
for its complicated shape on the way towards the dissociation limit. They constructed two 
PES versions based on extended ab initio calculations. Both PESs were computed at a high level of 
electronic structure theory with the largest basis sets ever used for ozone, $MRCI(+Q) / AVXZ$ with $X = 5, 6$ and 
extrapolation to the complete basis set limit.  The first PES, referred to as R\_PES (``reef\_PES'' ) 
has been obtained including a single electronic state in the orbital optimization. It  possesses the 
``reef'' TS feature, as most published potentials do. The second one accounts for Dawes {\it et al.}'s correction~\cite{DAW11:081102}
which considers interaction with excited states. This latter potential is referred to as NR\_PES 
(``no\_reef\_PES''). Both PESs have very similar equilibrium configurations in the bottom of the main 
$C_{2v}$ potential well and give the same dissociation threshold: the theoretical value for both of 
them, $D_0 = 1.0548 \, {\rm eV}  \approx 8508 \, \cm$,  lies between two experimental dissociation energies 
with a deviation of only 0.6\% from the most recent experimental value of Ruscic~\cite{RUS06:6592,ruscic2014updated}
(as cited in~\cite{HOL10:9927}). 
Vibrational calculations, using the NR\_PES by Tyuterev {\it et al.}~\cite{TYU13:134307}, of all $\rm (^{16}O)_3$ band centers observed in rotationally resolved spectroscopy experiments 
have resulted in an (RMS) obs-calc error of only $\sim  1 \, \cm$ 
without any empirical adjustment.

Metastable ozone states above the dissociation threshold are expected to play a key role in the two-step 
Linderman mechanism~\cite{BAB03:2577} of ozone formation at low pressures.  They have been studied by Babikov {\it et al.}~\cite{BAB03:2577}  
and by Grebenschikov and Schinke~\cite{GRE09:241,GRE09:181103} involving also lifetime calculations. Both investigations 
are based on SSB or mSSB potential surfaces~\cite{BAB03:2577} exhibiting the ``reef''-structure features. 
Assignment of recent very sensitive cavity-ring-down laser experiments in the TS energy range 
(from 70\% to 93\% of $D_0$) have been  possible~\cite{TYU14:143002,MON13:49,STA13:104,DEB13:24,BAR14:51,STA14:211,CAM15:84,STA15:203}
due to ro-vibrational predictions using the NR\_PES  
that changed the shape of the bottleneck range along the MEP and transformed the reef into a kind of smooth shoulder. 
The predictions of bound states with this latter PES in the TS energy range (from 70\% to 93\% of $D_0$) 
exhibit average errors of only $1 - 2 \, \cm$ for six ozone isotopologues, 666, 668, 686, 868, 886 and 888~\footnote{Here we
use a common abbreviation for the ozone isotopologues: $666 \equiv \rm (^{16}O)_3$, $668 \equiv \rm {^{16}O^{16}O^{18}O}$,
$686 \equiv \rm {^{16}O^{18}O^{16}O}$, etc.}. 
This clearly demonstrated~\cite{TYU14:143002} that the  NR\_PES by Tyuterev {\it et al.}~\cite{TYU13:134307} 
is much more accurate than other available surfaces for the description of all experimental spectroscopic data,
at least up to 8000~$\rm cm^{-1}$,
that is, for bound states up to at least 93\% of the dissociation threshold.  
In the original publication of Ref.~\cite{TYU13:134307}, bound states have been computed in the $C_{2v}$ symmetry of the main 
potential well.  To our knowledge no systematic studies of metastable ozone states with this NR\_PES~\cite{TYU13:134307}
have been published so far.

In the present work we report the first calculations of resonance state energies, corresponding wave functions 
and lifetimes using this PES. Furthermore, bound states near the dissociation threshold are investigated 
in full $D_{3h}$ symmetry, accounting for possible permutation of identical nuclei over the three potential wells.

\section{Symmetry considerations: Stationary approach}
\label{sec:theory}

In the electronic ground state, the ozone molecule has $C_{2v}$ symmetry at equilibrium such that the global 
potential energy surface has three relatively deep minima, corresponding to three possible arrangements of the oxygen atoms known as ``open configurations''.
As the barriers between two wells are very high, low-lying rovibrational states of the homonuclear ozone isotopologues,
such as $\rm (^{16}O)_3$, which we study in the present article,
may be characterized by irreducible representations (irreps) of the 
molecular symmetry group $C_{2v}(M)$, which is isomorphic with the $C_{2v}$ point group.
In the terminology of Longuet-Higgins~\cite{longuet1963symmetry,bunker98}, transformations between the three possible 
arrangements of three oxygen atoms in ozone are not feasible at low energies.

For weakly bound rovibrational states, however, for which tunneling of the barrier becomes noticeable,
and for continuum states of ozone above the barrier, the transformation between arrangements 
becomes feasible: The description of the dynamics of such states cannot be restricted to one potential well. 
In this situation, the complete molecular symmetry group must be employed to classify nuclear motion. 
This group is the three-particle permutation inversion group, 
$S_3 \times I$. It is isomorphic with the point group $D_{3h}$ and hence may also be designated 
$D_{3h}(M)$, where $M$ stands for molecular symmetry group~\cite{bunker98}.
Dissociation of the ozone molecule on the electronic ground state surface leads to an oxygen atom
and a dioxygen molecule, both in their electronic ground states, i.e. O($^3P$) + O$_2(X^3\Sigma_g^-)$.
The symmetry group of the oxygen atom is just the inversion group $I$, while that of the
oxygen molecule is the two-particle permutation inversion group $S_2 \times I$. The latter may
be designated $D_{\infty h}(M)$ in order to retain the $D_{\infty h}$ nomenclature for the irreducible
representations~\footnote{It is important here to note that 
no degenerate representations are included in the molecular symmetry group $D_{\infty h}(M)$,
in contrast to the point group $D_{\infty h}$,  which contains representations of type
$\Pi$, $\Delta$ etc. The order of the group $D_{\infty h}(M)$ is four, while that of the
point group $D_{\infty h}$ is infinite.}.
In the asymptotic channel, exchange of identical nuclei between the atom
and the diatomic molecule becomes unfeasible as their distance goes to infinity. It is clear
from this discussion that the molecular symmetry groups $C_{2v}(M)$ and $D_{\infty h}(M)$
are equivalent and just provide different sets of labels for the four irreducible representations.
They are two manifestations of the $S_2 \times I$ group.
To make this paper self-contained we give the characters and symmetry labels in Table~\ref{tab_C2v}. Of the symmetry elements of the point group $D_{\infty h}$ only those are
retained for the molecular symmetry group $D_{\infty h}(M)$ that correspond to a permutation inversion operation. This excludes
symmetry elements such as $2C(\phi)$ which leave all nuclei on their place.
The molecule is placed in the $xz$ plane, which is the convention  normally
used in ozone spectroscopy.~\footnote{This choice is different from that of Bunker \& Jensen~\cite{bunker98}.
As a result, the symmetry labels $B_1$ and $B_2$ are interchanged.}
The correspondence  of the axes
is thus ($x \rightarrow b, y \rightarrow c, z \rightarrow a$).
The transformation properties of the $p$ orbitals, which are needed in the discussion of the
asymptotic states, are indicated in the last column of the table. 
\begin{table}[t]
\caption{Character table of the point groups $C_{2v}$, $D_{\infty h}$ (excerpts) and the
permutation inversion group $S_2 \times I$  using the nomenclatures
of $C_{2v}(M)$ and $D_{\infty h}(M)$ for the irreducible representations.
}
\label{tab_C2v}
\begin{tabular}{cc|@{~~~}c@{~~~}c@{~~~}c@{~~~}c | c}
\hline \hline
\multicolumn{2}{c|@{~~~}}{$C_{2v}$} & E & $C_{2b}$ & $\sigma_{ab}$ &  $\sigma_{bc}$ & \\
\multicolumn{2}{c|@{~~~}}{$D_{\infty h}$} & E & $\infty C_2'$ & $\infty \sigma_v$ &  $i$ & \\ \hline 
\multicolumn{2}{c|@{~~~}}{$S_2 \times I$} & $E$ & (12) & $E^{*}$ &  $(12)^{*}$ & \\
$C_{2v}(M)$ & $D_{\infty h}(M)$    &   &      &         &            &   \\ \hline
$A_1$ &  $\Sigma_g^+$              & 1 &   1  &     1   &       1    &  $p_b$  ($p_x$)\\
$B_1$ &  $\Sigma_u^+$              & 1 &  -1  &     1   &      -1    &  $p_a$  ($p_z$)\\
$A_2$ &  $\Sigma_u^-$              & 1 &   1  &    -1   &      -1    &  \\
$B_2$ &  $\Sigma_g^-$              & 1 &  -1  &    -1   &       1    &  $p_c$ ($p_y$) \\
\hline \hline
\end{tabular}
\end{table}

Classification of states in $S_2 \times I$ is convenient for rovibrational states situated deep in the wells,
and for the dissociating resonances.  We now wish to relate them with the symmetry species of
the complete permutation inversion group $S_3 \times I$, or $D_{3h}(M)$. These correlations are
shown in Table~\ref{tab_D3h}. In addition to the symmetry elements of $S_2 \times I$,
which are the identity, $E$, the pair permutation, $(12)$, the inversion of the spatial coordinate
system, $E^{*}$, and the combination $(12)^{*} = (12) \times E^{*} = E^{*} \times (12)$, a
new class appears, the cyclic permutations, $\{(123), (132)\}$, as well as the class built up by
its combination with the inversion of the coordinate system, \{$(123)^{*}$, $(132)^{*}$\}.
These new operations describe the exchange between the three localized structures.
The correlation presented in Table~\ref{tab_D3h} is obtained by matching the characters of the
common operators, i.e. the identity operation, pair permutations and the inversion of the coordinate system.
\begin{table}[t]
\caption{Character table of the group $S_3 \times I$ and the relation with
the  irreducible representations of the group $S_2 \times I$ using the nomenclatures
$C_{2v}(M)$ and $D_{\infty h}(M)$.}
\label{tab_D3h}
\resizebox{\textwidth}{!}{%
\begin{tabular}{c|cccccc|c|c}
\hline \hline
$S_3 \times I$ & E & \{(123), (132)\} & \{(12), (23), (13)\} & $E^{*}$ &  \{$(123)^{*}$, $(132)^{*}$\} & \{$(12)^{*}$, $(23)^{*}$, $(13)^{*}$\}  & \multicolumn{2}{c}{$S_2 \times I$} \\
$D_{3h}(M)$    &   &    &     &    &    &      & $C_{2v}(M)$ & $D_{\infty h}(M)$ \\ \hline 
 $A'_1$        & 1 & ~1 & ~1  & ~1 & ~1 & ~1   & $A_1$ & $\Sigma_g^+$ \\
 $A'_2$        & 1 & ~1 & -1  & ~1 & ~1 & -1   & $B_1$ & $\Sigma_u^+$ \\
 $E'$          & 2 & -1 & ~0  & ~2 & -1 & ~0   & $A_1 + B_1$ & $\Sigma_g^+ + \Sigma_u^+$ \\
 $A''_1$       & 1 & ~1 & ~1  & -1 & -1 & -1   & $A_2$ & $\Sigma_u^-$ \\
 $A''_2$       & 1 & ~1 & -1  & -1 & -1 & ~1   & $B_2$ & $\Sigma_g^-$ \\
 $E''$         & 2 & -1 & ~0  & -2 & ~1 & ~0   & $A_2 + B_2$ & $\Sigma_u^- + \Sigma_g^-$ \\
\hline \hline
\end{tabular}
}
\end{table}

The rovibrational states of ozone may now be classified in the $D_{3h}(M)$ group, allowing for
tunneling between the three wells.  They can be considered  superpositions of the three states localized in their wells,
which give rise to a one-dimensional representation and a two-dimensional representation, just as in the case of 
triplet $\rm H_3^+$ which has been discussed before~\cite{alijah15}. The energy difference between the one and the two-dimensional
representations is called tunneling splitting. Purely vibrational states have
positive parity, i.e. belong to either $A'_1$, $A'_2$ or $E'$, while both prime and  double prime states exist 
for rotationally  excited states.
The localized vibrational states to be superimposed may be classified in $C_{2v}(M)$
by the approximate normal mode quantum numbers
$| v_1 \, v_2 \, v_3 \rangle$ of the symmetric stretching vibration, $v_1$, the bending vibration, $v_2$, and the antisymmetric stretching vibration, $v_3$.
Since these transform as $A_1$, $A_1$ and $B_1$, respectively, the symmetry of $| v_1 \, v_2 \, v_3 \rangle$
is $A_1$ for $v_3$ even and $B_1$ for $v_3$ odd. In $D_{3h}(M)$, they give rise to the pairs
($A'_1, E'$), ($A'_1, E'$) and ($A'_2, E'$), referring to the one and two-dimensional representations. 

Only those vibrational states that have $A'_1$ symmetry are allowed for the isotopologue $\rm (^{16}O)_3$ as can
be seen from the following analysis: 
The $^{16}$O isotope is a boson, with zero nuclear spin, i.e. the total wave function of $\rm (^{16}O)_3$ 
must be symmetric under exchange of any two $^{16}$O nuclei and transform as $A_1'$ or $A_1''$ in $D_{3h}(M)$. 
The nuclear spin function transforms as $A'_1$.
Likewise, the electronic wave function of the ground state, 
$X \, ^1A_1$ in spectroscopic notation, since the open structure minima have $C_{2v}$ symmetry, is totally symmetric with respect 
to all nuclear permutations. 
It means that the rovibrational part  of $(^{16}$O$)_3$ should also be symmetric under an exchange of any two oxygen nuclei, 
i.e. should transform as the $A_1'$ or the $A_1''$ irreducible representation. Purely vibrational states have positive parity
and thus symmetry $A_1'$, the other symmetry species are not allowed.
We note in particular that the degenerate tunneling component has zero statistical weight, giving rise to ``missing levels''
in spectroscopic language. As a consequence, tunneling splitting of the purely vibrational states cannot be observed.

The calculations of the present article were performed in hyperspherical coordinates, as they permit 
straightforward implementation of the full permutation inversion symmetry.
The rovibrational wave function $\Psi^{Jm}_{v}$ of tunneling ozone can be written as an expansion over products of 
rotational ${\cal R}_{Jkm}(\Omega)$ and vibrational factors $\psi_{v}^{Jk}({\cal Q})$
\begin{equation}
\label{eq:expansion}
\Psi^{Jm}_{v}(\Omega,{\cal Q})=\sum_{k} {\cal R}_{Jkm}(\Omega)\psi_{v}^{Jk}({\cal Q})\,,
\end{equation}
where ${\cal R}_{Jkm}(\Omega)$ are symmetric top rotational wave functions proportional to the Wigner functions $D^J_{mk}$
\begin{equation}
{\cal R}_{Jkm}(\Omega)=\sqrt \frac{2J+1}{8\pi^2}\left[D^J_{mk}(\Omega)\right]^*\,,
\end{equation}
and depending on the three Euler angles $\Omega$. The vibrational part of the wave function depends on the internal projection 
$k$ of the angular momentum onto the axis perpendicular to the molecular plane, denoted the $y$-axis in Table~\ref{tab_C2v}.
Note that no decomposition is made here in terms of the $C_{2v}(M)$ normal modes, which
would be an approximation.

Each product in expansion (\ref{eq:expansion}) should have the same symmetry in the $D_{3h}(M)$ group as the 
total rovibrational wave function, i.e. $A_1'$ or $A_1''$. The symmetry 
$\Gamma^{r}$ of the rotational functions ${\cal R}_{Jkm}(\Omega)$ in $D_{3h}(M)$ is well known
(see, for example, \cite{bunker98,kokoouline03b}). It imposes restrictions on the possible irreducible representations 
of the vibrational factors $\psi_{v}^{Jk}({\cal Q})$: The rotational and vibrational wave functions should be of the same species, 
both $A_1$, or both $A_2$, or both $E$. Parities of the wave functions are not restricted. The parity of the 
vibrational functions is always positive, the parity of the rotational function is positive for even $k$ and 
negative for odd $k$. Examples of the irreducible representations of rotational and vibrational functions 
are given in Table \ref{tab:irreps} for $J \le 3$.

\begin{table}
\caption{Allowed combinations of irreducible representations of the rotational and vibrational factors in the expansion of 
Eq.~(\ref{eq:expansion}). \footnote{The symmetrized combinations of functions with $k=\pm 3$ transform as the $A_1''$ and $A_2''$ 
representations in $D_{3h}(M)$. The direct products, $\Gamma^r \times \Gamma^v$, of two $E$ representations
yield $A_1 + A_2 + E$ and contain the $A_1$ representation. Only the latter is listed in the last line of the table.
The parity is given by the usual rule $' \times '= '$, $'' \times ''= '$, $' \times ''= ''$, $'' \times '= ''$.}
}
\label{tab:irreps}
\begin{tabular}{c p{1cm} p{1cm} p{1cm} p{1cm} p{1cm} p{1cm} p{1cm} p{1cm} p{1cm} p{1.3cm} }
\hline \hline
$J$ 		&0	&1		&1 &2 &2&2&3&3&3&3 \\
\hline
$k$ 		&0	&0		&$\pm$1 &0 &$\pm$1&$\pm$2&0&$\pm$1&$\pm$2&$\pm$3$^*$\\
$\Gamma^r$	&$A_1'$	&$A_2'$		& $E''$&$A_1'$&$E''$&$E'$&$A_2'$&$E''$&$E'$&$A_1''$, $A_2''$\\
$\Gamma^v$	&$A_1'$	&$A_2'$		& $E'$&$A_1'$&$E''$&$E'$&$A_2'$&$E'$&$E'$&$A_1'$, $A_2'$\\ \hline
$\Gamma^r \times \Gamma^v$ & $A_1'$ & $A_1'$ & $A_1''$ & $A_1'$ & $A_1'$ &  $A_1'$ & $A_1'$ &  $A_1''$ & $A_1'$ & $A _1''$, $A _1''$ \\
\hline
\hline
\end{tabular}
\end{table}

Let us now turn to the symmetry classification of the wave functions of the decaying resonance states.
The lowest dissociation limit of ozone produces the oxygen atom, O$\, (^3P)$, 
and the oxygen molecule, O$_2\, (X^3\Sigma_g^-)$, in  their electronic ground states.
The orbital degeneracy of the atomic $P$ state is three. One orbital is oriented perpendicular to the
plane spanned by the three nuclei, denoted as $p_c$ in Table~\ref{tab_C2v}. According to Table~\ref{tab_D3h},
it transforms as $A_2''$ in $D_{3h}(M)$. The two in-plane orbitals transform as $E'$. On the other hand,
the electronic symmetry of the di-oxygen molecule is $\Sigma_g^-$ in $D_{\infty h}(M)$, or $A_2''$ in $D_{3h}(M)$.
At large distances, the electronic ground state, $X \, ^1A_1$, of ozone correlates with the
perpendicular ($p_c$) component of the atomic $P$ state plus the diatomic $\Sigma_g^-$ state, which have both 
$A_2''$ symmetry in $D_{3h}(M)$ such that their product is indeed $A_1'$.

The electronic ground state of $\rm O_2$ is antisymmetric with respect to an exchange of the two nuclei.
Since the vibrational states of $\rm O_2$ are totally symmetric,
this implies that the rotational functions must be antisymmetric to yield a symmetric nuclear 
wave function. The rotational functions of $\rm O_2$ transform as $\Sigma_g^+$ for even values of $j$ and
as $\Sigma_g^-$ for odd values. Rotational states of $\rm ^{16}O_2$ must therefore have 
odd rotational angular momentum, $j$,
and the lowest rovibrational state is ($v=0, j=1$). 

Let us now analyze the asymptotic wave function in the exit channel $\kappa$ with $\kappa = 1, 2, 3$. 
It can be expanded as
\begin{equation}
\label{eq:scattering}
\Psi^{Jm}_{\kappa v_d j l}(\vec{r}_{\kappa}, \vec{R}_{\kappa}) \approx \frac{1}{r_{\kappa} R_{\kappa}} \varphi^{el}_a\varphi^{el}_d
\chi_{v_dj}(r_{\kappa}) {\cal Y}^{Jm}_{jl}(\hat{r}_{\kappa}, \hat{R}_{\kappa})  e^{i (k R_{\kappa} - l \pi/2)}\,,
\end{equation}
where   $\exp(i (k R_{\kappa} - l \pi/2))$ is the scattering function of the outgoing wave 
and $\chi_{v_dj}(r_{\kappa})$ the vibrational wave function of the $\rm O_2$ molecule; 
$r_{\kappa}$ and ${R}_{\kappa}$ are the true, not mass-scaled, distances in the Jacobi coordinate system $\kappa$. Functions $\varphi^{el}_a$ and $\varphi^{el}_d$ represent electronic states of the O$\,(P)$ atom and the O$_2\,(X^3\Sigma_g^+)$ molecule.
Angular momenta of the atom-diatom relative motion, $l$, and of the rotation of the oxygen molecule, $j$, must be coupled
to yield the total angular momentum, $J$, which is taken care of by the bipolar harmonics, ${\cal Y}^{Jm}_{jl}$.
They are defined as
\begin{equation}
{\cal Y}^{Jm}_{jl}(\hat{r}_{\kappa}, \hat{R}_{\kappa}) = \sum_{m_l, m_j} C_{j m_j l m_l}^{Jm} Y_{j m_j}(\hat{r}_{\kappa}) Y_{l m_l}(\hat{R}_{\kappa})\,,
\end{equation}
where the $Y$ are spherical harmonics and $C$  are Clebsch-Gordan coefficients. The scattering function in Eq.~(\ref{eq:scattering}) is not symmetric with respect to permutation of three bosonic nuclei and, therefore, cannot be correlated in this form with the short-distance form of Eq.~(\ref{eq:expansion}), which does have correct symmetry behavior (for the combinations of quantum numbers  given in Table \ref{tab:irreps}). To bring the function of Eq.~(\ref{eq:scattering}) to the form satisfying the permutationl symmetry of three bosons, in the language of group theory, one has to apply projectors of the $D_{3h}(M)$ group of the two allowed irreducible representations, $A_1'$ or $A_1''$. An efficient way to perform it is to use a general approach  of Ref.~\cite{douguet08a} applicable to a three-body system with arbitrary total nuclear spin. Equations (19) of that reference do not take into account the electronic part of the total wave function. The electronic wave function of the dioxygen $\varphi^{el}_d$ changes sign under permutation of the two atoms and under the inversion operation, and the atomic $\varphi^{el}_a$ changes sign under the inversion only. Therefore, Eqs.~(19) of Ref.~\cite{douguet08a} take the following form for the present case
\begin{eqnarray}
 \label{eq:12}
(12) \Psi^{Jm}_{\kappa v_d j l}(\vec{r}_{\kappa}, \vec{R}_{\kappa}) =(-1)^{j+1}\Psi^{Jm}_{\kappa v_d j l}(\vec{r}_{\kappa}, \vec{R}_{\kappa})\nonumber\\
E^* \Psi^{Jm}_{\kappa v_d j l}(\vec{r}_{\kappa}, \vec{R}_{\kappa}) =(-1)^{l+j}\Psi^{Jm}_{\kappa v_d j l}(\vec{r}_{\kappa}, \vec{R}_{\kappa})
\end{eqnarray}
With these properties, the projectors $P_{\Gamma}$ take the form (see Eqs.~(20) of Ref.~\cite{douguet08a})
\begin{eqnarray}
\label{eq:projector_21adapted2}
P_{\Gamma}\Psi^{Jm}_{\kappa v_d j l}(\vec{r}_{\kappa}, \vec{R}_{\kappa})=\nonumber\\
\left(1+ \chi^{\Gamma}_{23}(23)+\chi^{\Gamma}_{31}(31)\right) \left(1 + (-1)^{j+1}\chi^{\Gamma}_{12}\right)\left(1+ (-1)^{l+j}\chi^{\Gamma}_{E^*}\right)\Psi^{Jm}_{\kappa v_d j l}(\vec{r}_{\kappa}, \vec{R}_{\kappa})\,,
\end{eqnarray}
for any of the $D_{3h}(M)$ representations.
Here, $\chi^{\Gamma}$ are characters of the representation $\Gamma$ given in Table~\ref{tab_D3h}. From the expression in the second parentheses on the right side of the equation above, it is clear that for the allowed representations $A_1'$ and $A_1''$, if $j$ is even, the projectors are identically zero, ${P}_{A_1'} = 0$, ${P}_{A_1''} = 0$. It is simply means that a free molecule $^{16}$O$_2 \, (X^3\Sigma_g^+)$ can only have odd rotational angular momentum $j$. The expression in the  third parentheses means that if the quantum numbers $l$ and $j$ have different parity, the projectors again give identically zero for $A_1'$ (but not for $A_1''$).  In particular, it implies that dissociative states of $^{16}$O$_3$ with rotational angular momentum $J=0$ do not exist within the adiabatic approximation.



\section{Nuclear dynamics}

The present, stationary theoretical approach to describe nuclear dynamics was developed previously by
Kokoouline {\it et al.}~\cite{kokoouline06,blandon07,blandon09,alijah15}. It is based on  the two-step procedure of solving the stationary Schr\"odinger equation in hyperspherical coordinates~\cite{johnson80,johnson83a,johnson83b}. Although the method was previously applied to several three-body problems, it has never been applied to a system with large masses of the three particles and so many bound states: In Ref.~\cite{blandon07} the method was developed and tested on a benchmark system of a three-boson nucleus with a very shallow potential supporting only one bound state and one resonance. In~\cite{blandon09}, the method was employed to calculate resonances in three-body collisions of hydrogen atoms. The lowest H$_3$ potential energy surface has two coupled sheets without any bound state but with many resonances. The method was also routinely used to represent the vibrational continuum in studies of dissociate recombination of isotopologues of H$_3^+$~\cite{kokoouline04a,santos07}. An important difference of the present study with the previous ones is that the number of bound states is large, which requires a significantly larger basis to represent the vibrational dynamics near and above the dissociation.

We briefly summarize the main elements of the approach. To solve the  Schr\"odinger equation 
\begin{equation}
\label{eq:Schr1}
\left[T(\rho,\theta,\phi)+V(\rho,\theta,\phi)\right]\Phi_v(\rho,\theta,\phi)=E_v\Phi_v(\rho,\theta,\phi)
\end{equation}
for three particles interacting through the potential $V(\rho,\theta,\phi)$ in the hyperspherical coordinates $\rho,\theta$, and $\phi$, first, the adiabatic hyperspherical curves $U_a(\rho)$ and the corresponding hyperangular eigenstates $\varphi_a(\rho_i;\theta,\phi)$ (hyperspherical adiabatic states -- HSA) are obtained by solving the equation in the two-dimensional space of the hyperangles $\theta$ and $\phi$ for several fixed values of the hyper-radius $\rho_j\ (j=1,2,\cdots)$, i.e. the following equation is solved
\begin{equation}
\label{eq:Had}
\left[\hbar^2\frac{\Lambda^2+\frac{15}{4}}{2\mu \rho_j^2}+V(\rho_i;\theta,\phi)\right] \varphi_a(\rho_j;\theta,\phi)=U_a(\rho_i)\varphi_a(\rho_j;\theta,\phi).
\end{equation} 
In the above equation, $\Lambda^2$ is  the grand angular momentum squared~\cite{WHI68:1103,johnson83b} and $\mu$ is the three-particle reduced mass: For identical oxygen atoms with mass $m_O$, one has $\mu=m_O/\sqrt{3}$. The equation is solved using the approach described in~\cite{esry97}. Solution of Eq.~(\ref{eq:Had}) yields  adiabatic curves $U_a(\rho)$ and  eigenfunctions  $\varphi_a(\rho;\theta,\phi)$, defining a set of HSA channels $a$. The HSA states are then used to expand the wave function  $\Phi_v$ in Eq.~(\ref{eq:Schr1})
\begin{equation}
\label{eq:vibr_func_SVD}
\Phi_{v}({\cal Q})=\sum_a  \psi_a(\rho_j) \varphi_a(\rho_j;\theta,\phi)\,.
\end{equation}
 The  expansion coefficients $\psi_a(\rho_i)$ depend on hyper-radius $\rho$. Following the original idea of Ref. \cite{tolstikhin96} the hyper-radial wave functions $\psi_a(\rho_i)$ are then expanded in the discrete variable representation (DVR) basis $\pi_j(\rho)$
\begin{equation}
\psi_a(\rho)=\sum_j  c_{j,a}\pi_j(\rho).
\end{equation}
Inserting the two above expansions into the initial Schr\"odinger equation  (\ref{eq:Schr1}), one obtains
\begin{equation}
\label{eq:gener_eigen2}
\sum_{j',a'}\left[\langle\pi_{j'} |-\frac{\hbar^2}{2\mu}\frac{d^2}{d\rho^2}|\pi_j\rangle{\cal O}_{j'a',ja}+U_a(\rho_j)\delta_{j',j}\delta_{a'a}\right]c_{j'a'}=E\sum_{a'}{\cal O}_{ja',ja}c_{ja'}\,
\end{equation}
with
 \begin{equation}
{\cal O}_{j'a',ja}=\langle\varphi_{a'}(\rho_{j'};\theta,\phi)|\varphi_a(\rho_j;\theta,\phi)\rangle .
\end{equation}
In the above equation, the matrix elements of the second-oder derivative with respect to $\rho$ is calculated analytically (see, for example, \cite{tuvi1997hermiticity,kokoouline99} and references therein).

The described approach of solving the Schr\"odinger equation using the adiabatic (HSA) basis replaces the usual form of non-adiabatic couplings in terms of derivatives with respect to  $\rho$ with overlaps between adiabatic states $\varphi_a(\rho,\theta,\phi)$ evaluated at different values of $\rho$. 
The approach is particularly advantageous here, since the adiabaticity of the hyper-radial motion, 
when separated from hyperangular  motion, is not satisfactory, 
so that multiple avoided crossings between HSA energies $U_a(\rho)$ occur. 
This is the usual situation in three-body dynamics. 
Representing non-adiabatic couplings by  derivatives $\langle\varphi_{a'}\lvert \partial/\partial\rho\rvert\varphi_{a}\rangle$ and $\langle\varphi_{a'}\lvert \partial^2/\partial\rho^2\rvert\varphi_{a}\rangle$ near the avoided crossings would require a very small grid step in $\rho$. The use of overlaps between HSA states reduces significantly the number of grid points along $\rho$ required for accurate representation of vibrational dynamics.

\begin{figure}[ht]
\vspace{2cm}
\includegraphics[width=15cm]{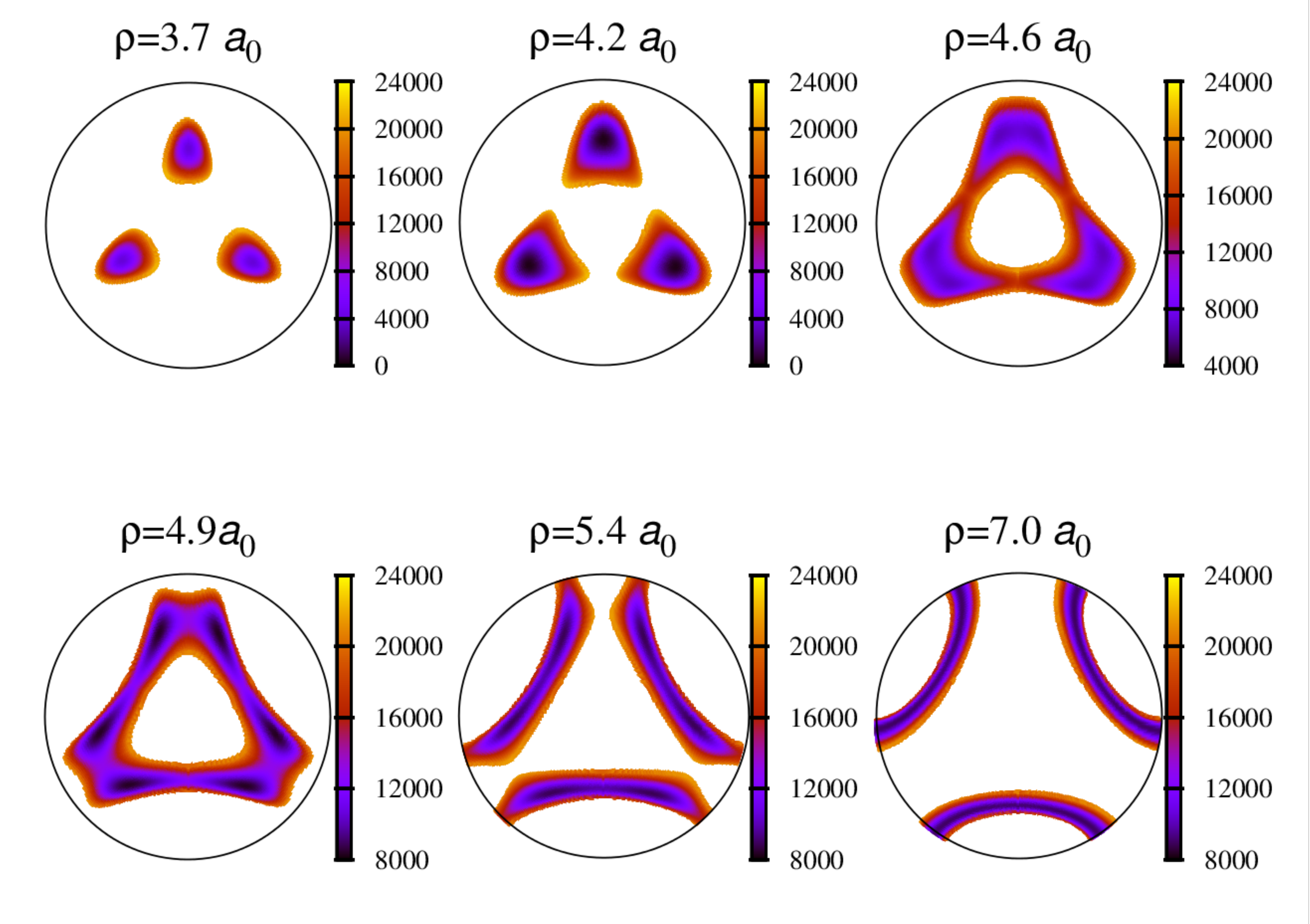}
\caption{Ozone potential energy surface, NR\_PES of Ref.~\cite{TYU13:134307}, as a function of the two hyperangles for several values of the hyper-radius. In the plots, the hyperangles are represented in a polar coordinate system (see Fig.~6 of Ref.~\cite{kokoouline03b}): $\theta$ increases from the center of each plot to its edge; $\phi$ is a cyclic variable (polar angle) changing from 0 to $2\pi$. The minimum of PES, situated near $\rho=4.2~a_0$, is chosen as origin. The electronic energy of  dissociation to the atom and the diatomic molecule at equilibrium is at 9150 cm$^{-1}$.}
\label{fig:pes}
\end{figure}

In Ref.~\cite{TYU13:134307}, the main features of the PES were demonstrated in internal coordinates. 
In the present study, the NR\_PES of Ref.~\cite{TYU13:134307}, which had been originally defined 
in the $C_{2v}$ wells, was symmetrized according to the nuclear permutations and converted in the 
hyperspherical coordinates~\cite{johnson80,johnson83a,johnson83b}.
Fig.~\ref{fig:pes} shows the PES as a function of the two hyper-angles for several values of the hyper-radius. As evident from the plot at $\rho=5.4$ bohr the potential barrier between the wells is situated at energies 9000 cm$^{-1}$, i.e. very close to the dissociation threshold. The passage between the wells occurs at geometries beyond the ``shoulder'' of the ozone potential. Therefore, one expects weakly bound low-energy resonances delocalized between the three potential wells. To represent nuclear dynamics of such  near-dissociation levels, one needs to take into account the three potential wells simultaneously.  
The energy  $D_0$ of dissociation to the  ${\rm ^{16}O} \, (^3P)$ and 
${\rm ^{16}O_2} \, (X \,{^3\Sigma}_g^- [v_d = 0, j = 0])$ products is 8555 cm$^{-1}$ above the ground rovibrational level of $\rm ^{16}O_3$.

\begin{figure}[ht]
\includegraphics[width=12cm]{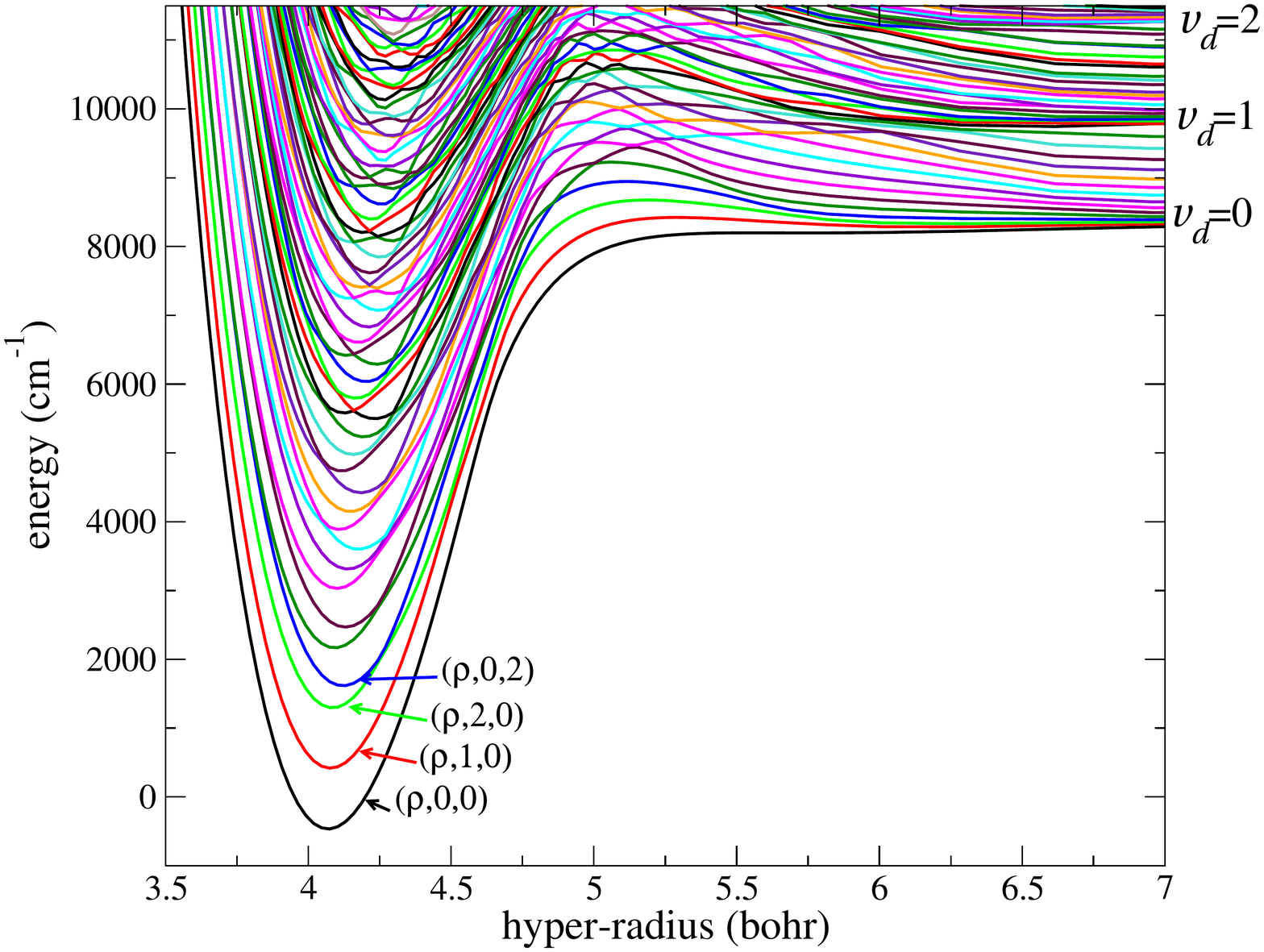}
\caption{Hyperspherical adiabatic curves $U_a(\rho)$ of the $A_1$ irreducible representation and $J=0$ as a function of hyper-radius obtained for $^{16}$O$_3$. In this figure, the energy origin is chosen at the ground vibrational level of $^{16}$O$_3$, situated 1443.524 cm$^{-1}$ above the PES minimum. The $v_d=0,1,2$ labels indicate energies of the ${\rm O \,  + \, O_2} \, (v_d)$ asymptotic vibrational channels. Multiple HSA curves between the vibrational channels correspond to various rotational channels $j$ of the dissociating oxygen molecule. For the $A_1$ vibrational states, only even $j$ are allowed. Because $J=0$, the partial wave in each asymptotic channel $(v_d,j,l)$ is determined simply as $l=j$.}
\label{fig:adiabatic}
\end{figure}

A convenient way of analyzing nuclear dynamics of three atoms is given by HSA curves, which could be viewed in a way similar to Born-Oppenheimer curves for diatomic molecules, except that the adiabatic and dissociation coordinate in the HSA curves is the hyper-radius, not the inter-atomic distance. In contrast to the case of Born-Oppenheimer separation between electronic and vibrational motion for diatomic molecules, non-adiabatic coupling between HSA states is almost always strong and cannot be neglected. Nevertheless, many key features of the dynamics can easily be identified and qualitatively studied. The HSA curves obtained for $A_1$ vibrational symmetry and $J=0$ are shown in Fig.~\ref{fig:adiabatic}. At small values of hyper-radius, near $\rho=4.2$, the lowest HSA curves have a minimum, which corresponds to the O$_3$ equilibrium. Each of the lowest HSA curves near the minimum represents approximately a particular combination of $v_2$ and $v_3$ vibrational modes of  O$_3$. The $v_1$ mode near the O$_3$ equilibrium is represented by the continuous variable $\rho$, which is at this first step not quantized in the space of HSA coordinates. Therefore, the lowest HSA curve $U_a(\rho)$ ($a=1$) near $\rho=4.2$ is an adiabatic representation of the set $(\rho,0,0)$ of vibrational modes of O$_3$ corresponding to the normal mode quantum numbers $v_2 = v_3 = 0$, the second and third HSA curves are $(v_2 = 1, v_3 = 0)$ and $(v_2 = 2, v_3 = 0)$, the fourth one is $(v_2 = 0, v_3 = 2)$, etc. Odd $v_3$ are not present in $A_1$ vibrational symmetry.

At energies near and higher than 6000 \cm above the $(0,0,0)$ level, the normal modes are significantly mixed and the mode assignment becomes more difficult. However, the HSA curves at large energies, above the energy of dissociation, and at large $\rho$, provide a convenient description of dissociation dynamics. At large $\rho$, each adiabatic curve converges to a particular asymptotic channel represented by a rovibrational level $(v_d,j)$ of  O$_2$  and the partial wave of relative motion of O$_2$ and O. As one can see, there are multiple very sharp avoided crossings, especially in the zone of transition from short to large $\rho$.

\section{Bound states near $\bm{D_0}$ and predissociated resonances}

\begin{figure}[ht]
\vspace{1cm}
\includegraphics[width=13cm]{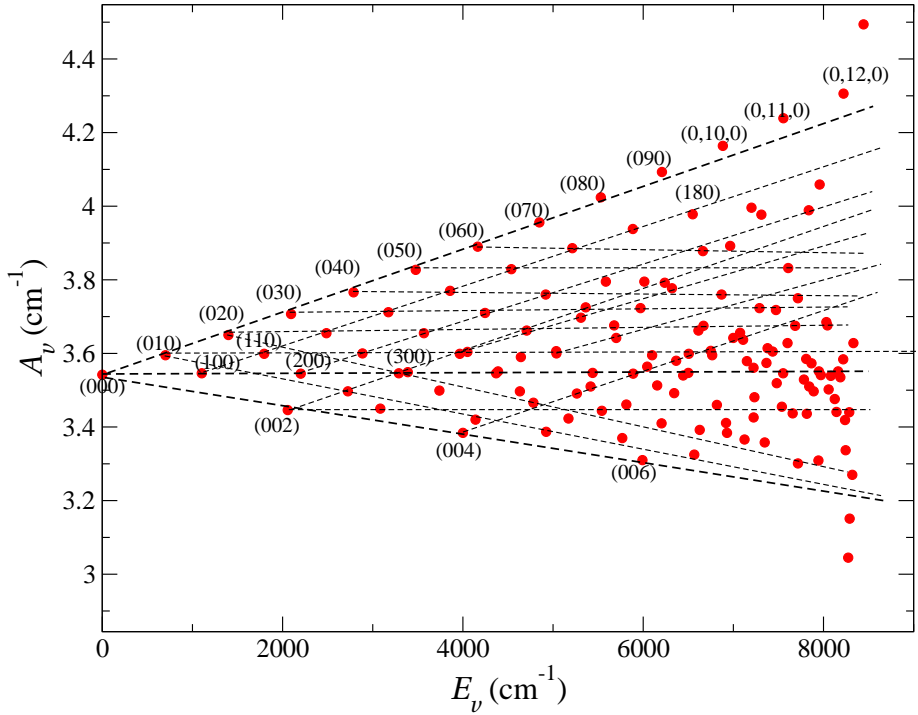}
\caption{The largest of the three rotational constants for $A_1$ vibrational states in the $D_{3h}$ group. Almost linear dependence of the rotational constant $A_v$ on the energy of the vibrational states permits an assignment of normal modes for low energy levels. Normal mode quantum numbers are specified for a few levels.
Note than at high energy the normal mode assignment becomes ``nominative'' and is to be taken with 
caution because of strong anharmonic basis state mixing.}
\label{fig:Bv_A1}
\end{figure}

\begin{figure}[ht]
\vspace{1cm}
\includegraphics[width=11cm]{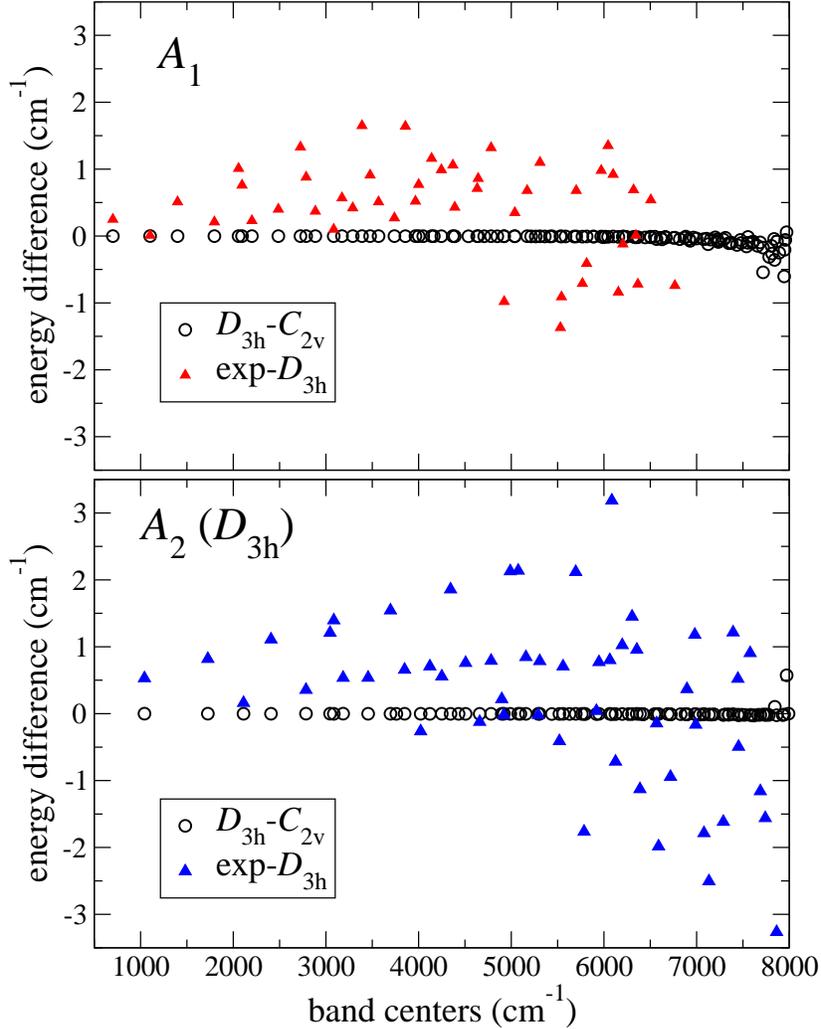}
\caption{Comparison of the energies of band centers obtained  in this study  with the previous calculation \cite{TYU13:134307} and experimental data  \cite{TYU14:143002,CAM06:1,BAR13:172,MON12:840,BAB14:169,MON13:49,STA13:104,DEB13:24,BAR14:51,STA14:211,CAM15:84} for two vibrational symmetries ($A_1$ and $A_2$ in the $D_{3h}$ group employed here, $A_1$ and $B_1$ in the $C_{2v}$ group employed in Ref.~\cite{TYU13:134307}). The difference between the present results and the previous calculation and experimental data is labeled as $D_{3h}-C_{2v}$ and exp$-D_{3h}$ respectively.}
\label{fig:theory_exp}
\end{figure}

A series of calculations with different parameters of the numerical approach were performed to assess the uncertainty of the obtained energies with respect to the numerical procedure. The final results for $A_1$ and $A_2$ vibrational levels were obtained with 60 HSA states. The number of $B$-splines used for each of the hyperspherical angles $\theta$ and $\phi$ was 120. Similar to previous work by Alijah and Kokoouline
on the $\rm H_3^+$ molecule~\cite{alijah15}, the interval of variation of $\varphi$ was from $\pi/6$ to $\pi/2$ in calculations of $A_1$ and $A_2$ levels. The variation interval of $\rho$ was from $2.9$ to $16$, a variable step width \cite{kokoouline99,kokoouline06,blandon09} along the $\rho$ grid was used with 192 grid points. The estimated uncertainty due to the employed numerical method is better than $0.001$~\cm for low vibrational levels and about 0.01~\cm for levels at around 7500~\cm above the ground vibrational level.
This convergence error is significantly lower than the uncertainties of the ozone PES. 
Figure~\ref{fig:theory_exp}  compares the energies of $^{16}$O$_3$ band centers up to 8000~\cm obtained  in this study  with the previous calculation \cite{TYU13:134307} and experimental data  \cite{TYU14:143002,CAM06:1,BAR13:172,MON12:840,BAB14:169,MON13:49,STA13:104,DEB13:24,BAR14:51,STA14:211,CAM15:84}. The RMS deviation between the calculation of  Ref.~\cite{TYU13:134307} in $C_{2v}$ symmetry  and the present $D_{3h}$ calculations is of 0.03~\cm only up to this energy cut-off. This confirms a good nuclear basis set convergence of both methods.  The RMS (obs.-calc.) deviation for all vibrational band centers directly observed in high-resolution spectroscopy experiments is 1~\cm. This is by one order of magnitude better than the accuracy of vibrational calculations using other ozone PESs available in the literature. The uncertainty in the determination of resonance energies depends on their widths and is roughly 10\% of the respective width. The uncertainty in calculated widths is better than 20\% for most of the resonances.

The assignment of vibrational bands is simplified by using the vibrational dependence of rotational 
constants predicted from the PES and derived from ro-vibrational spectra analyses as described 
in Refs.~\cite{BAR13:172,MON13:49,STA13:104,DEB13:24,BAR14:51,STA14:211,STA15:203}.
The largest rotational constant, $A_v$, corresponding to the ``linearization'' $z$-axis, 
is given by the following expression in hyper-spherical coordinates \cite{kokoouline04b}
\begin{eqnarray}
 A_v=\langle\Psi^{00}_{v} \lvert \frac{1}{\mu\rho^2(1-\sin\theta)}\rvert\Psi^{00}_{v}\rangle\,.
\end{eqnarray}
At low vibrational excitations, the rotational constant $A_v$ has nearly linear behavior with respect to the normal mode quantum numbers, $v_1, v_2$, or $v_3$, with proportionality coefficients different for each mode. This can be seen in Fig.~\ref{fig:Bv_A1}. For example, when $v_1=v_3=0$, the levels $(0,v_2,0)$ form almost a straight line in the $A_v(E_v)$ plot. The same is true for other series, $(v_1,0,0)$, $(0,0,v_3)$, $(1,v_2,0)$, etc. Near the dissociation limits, the normal mode approximation is not valid any more and the series become mixed, although, the $(v_1,0,0)$ and $(0,v_2,0)$ series survive even above the dissociation. 
Such states cannot dissociate into $\rm O_2 \, + \, O$ unless mixed with the antisymmetric vibrational mode.

\begin{figure}[ht]
\vspace{1cm}
\includegraphics[width=12cm]{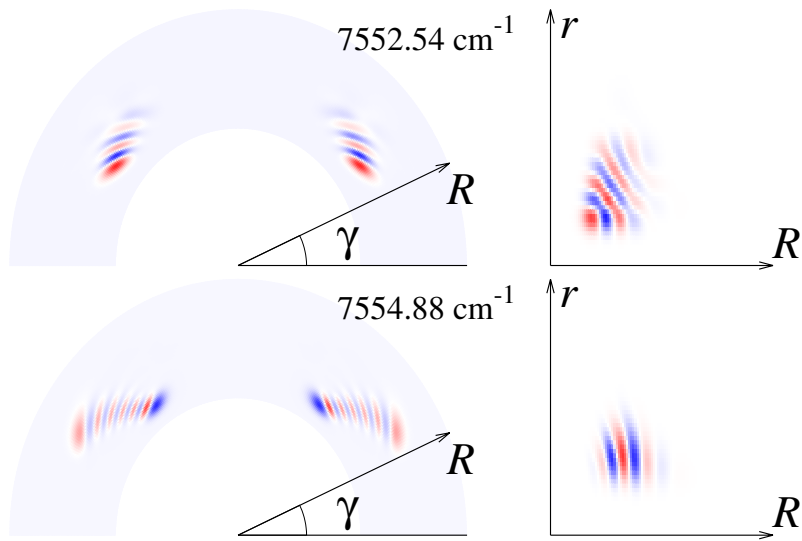}
\caption{Wave functions of the  $(7,0,0)$  (upper panels) and the $(0,11,0)$ (lower panels) levels  as functions of the Jacobi coordinates $R, r,$ and $\gamma$. On the left panels, the dependence on $R$ and $\gamma$ is shown  for a fixed value of $r=2.28\ a_0$, which is the equilibrium
nuclear distance in the $\rm O_2$ molecule. The right panels show the $R, r$-dependence for fixed $\gamma=40^\circ$. }
\label{fig:Jacobi_A1_v1_v2}
\end{figure}

\begin{figure}[ht]
\vspace{1cm}
\includegraphics[width=12cm]{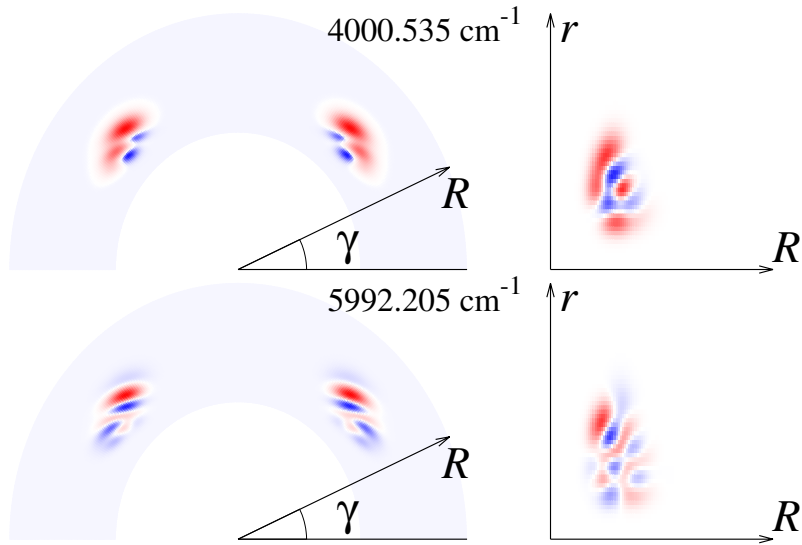}
\caption{As Fig.~\ref{fig:Jacobi_A1_v1_v2}, but for the pure antisymmetric stretching modes $(0,0,4)$ (upper panel) and $(0,0,6)$  (lower panel).}
\label{fig:Jacobi_A1_v3}
\end{figure}

\begin{figure}[ht]
\vspace{1cm}
\includegraphics[width=12cm]{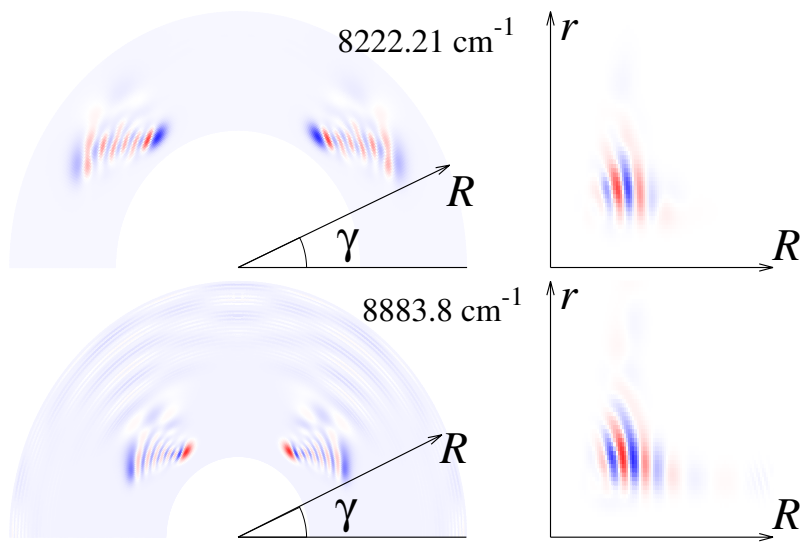}
\caption{As Fig.~\ref{fig:Jacobi_A1_v1_v2}, but for the $(0,12,0)$ bound (upper panel)  and $(0,13,0)$ resonance (lower panel) vibrational states. The interval or variation of $R$ in the left panel plots is larger than in  Figs.~\ref{fig:Jacobi_A1_v1_v2} and \ref{fig:Jacobi_A1_v3} in order to demonstrate the long-range tail of the wave functions. Note that the state shown in the lower figure exists only if rotation is excited. Thus, the wave functions are just the vibrational parts of the total rovibrational functions.
}
\label{fig:Jacobi_A1_v2_diss}
\end{figure}

Figures~\ref{fig:Jacobi_A1_v1_v2}, \ref{fig:Jacobi_A1_v3} and the upper panel of Fig.~\ref{fig:Jacobi_A1_v2_diss} show wave functions of five bound vibrational levels of $A_1$ vibrational symmetry in terms of Jacobi coordinates $R, r$, and $\gamma$, where $r$ is the distance between two oxygen nuclei of a chosen pair, $R$ is the distance from the center of mass of this pair to the third nucleus, and $\gamma$ is the angle between the vectors along $r$ and $R$. The left panels of the figures demonstrate the dependence of the wave functions on $R$ and $\gamma$. The interval of variation of $\gamma$ is from $0^\circ$ to $180^\circ$, such that it covers two of the three possible equivalent arrangements (permutations) of the three nuclei, i.e. it represents two of the three potential wells of the ozone potential. As evident, the obtained wave functions are symmetric with respect to an exchange between the two wells. Since the calculations were performed in hyper-spherical coordinates, the wave functions are also symmetric with respect to the exchange involving the third well, but the Jacobi coordinates cannot easily represent such a symmetry.

To demonstrate the nature of wave functions of different normal modes, the functions chosen in Figs.~\ref{fig:Jacobi_A1_v1_v2}, \ref{fig:Jacobi_A1_v3}, and \ref{fig:Jacobi_A1_v2_diss} represent ``pure'' vibrational modes:  $(7,0,0)$, $(0,11,0)$, $(0,12,0)$, $(0,0,4)$, and $(0,0,6)$. It is easy to identify the pure $v_1$ (symmetric stretching) and $v_2$ (bending) modes by counting  nodes in the Jacobi coordinates, but the behavior of the antisymmetric stretching mode $v_3$ is more complicated in Jacobi coordinates.

For the calculation of states above the dissociation threshold $D_0$, a complex absorbing potential (CAP) and variable grid step along $\rho$ adapted to the local de~Broglie wave length were used as described in Ref.~\cite{blandon07}. The parameters of the CAP were chosen to absorb the outgoing dissociation flux for the interval of energies approximately between 100 and 4000 \cm. When the method of CAP is used, the spectrum of the Hamiltonian matrix for energies above the dissociation limit contains not only the relatively long living resonance states but also non-physical ``box states''. Real and imaginary parts of box state eigenvalues depend on the CAP and grid parameters. A manual separation of resonances and box states is difficult for this case because of a large number of resonances. Several calculations with variable parameters, such as CAP, the number of grid pints along $\rho$, the number of the HSA states, the number of $B$-splines in the HSA calculations, were performed. Spectra obtained with different sets of parameters were compared, allowing us to separate the box-states from the resonances, as the latter no not
depend on the numerical parameters in a converged calculation.

The lower panel of Fig.~\ref{fig:Jacobi_A1_v2_diss} gives an example of a resonance wave function of $A_1$ vibrational symmetry. As discussed above, such levels are not allowed for $^{16}$O$_3$, but we will consider them because the same analysis can be applied to other isotopologues of O$_3$,
and also because a similar behavior can be exhibited by $A_1$ vibrational factors of rotationally excited $A_2$ states which are allowed for $^{16}$O$_3$. At short distances, the resonance is mainly described by the $(0,13,0)$ normal mode contribution. Its wave function looks very similar to that of the $(0,12,0)$ level.  It is still bound but has one more node along the $v_2$ coordinate. 
The outgoing dissociative flux is clearly visible in the $R,\gamma$ plot. The contrast in the $R,r$ plot is not quite sufficient to see the flux clearly. 
The vibrational resonance $(0,13,0)$ corresponds to large-amplitude bending motion of ozone. The energy of such bending oscillations is above dissociation, but the system does not dissociate fast, because the O$_2 \, + \, O$ dissociation implies that two of the three internuclear distances should become very large and the third distance should stay small, whereas when the molecule oscillates in  the $v_2$ or the $v_1$ modes, all three internuclear distances increase simultaneously.

\begin{figure}[ht]
\vspace{1cm}
\includegraphics[width=13cm]{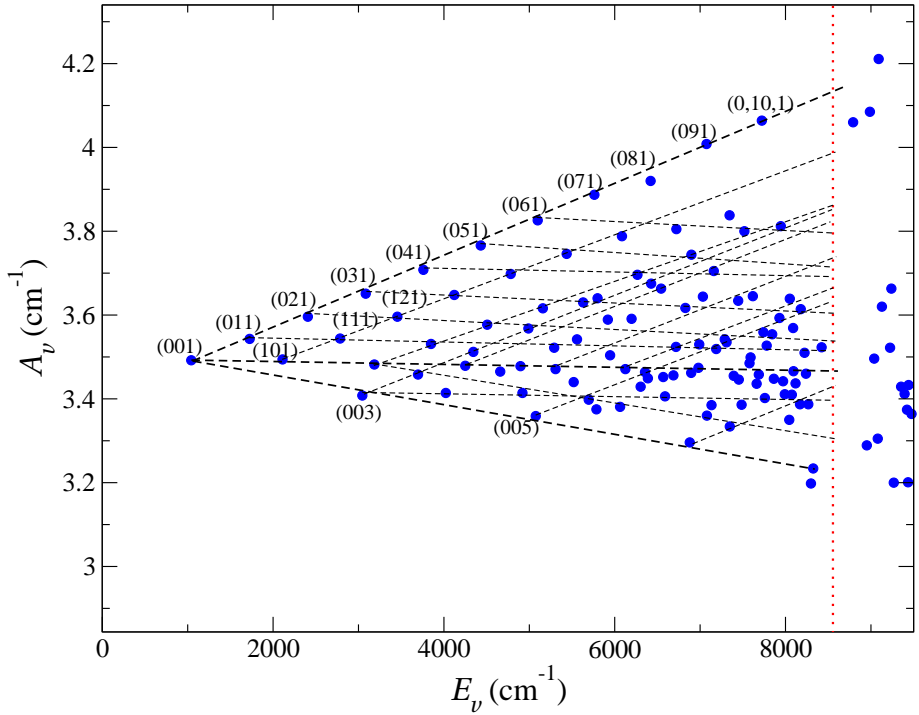}
\caption{The $A_v$ rotational constants for $A_2 (D_{3h})$ vibrational states.  Normal mode quantum numbers are specified for a few levels. Above the dissociation threshold (vertical dotted red line), the vibrational levels are predissociated.}
\label{fig:Bv_A2}
\end{figure}

\begin{figure}[ht]
\vspace{1cm}
\includegraphics[width=12cm]{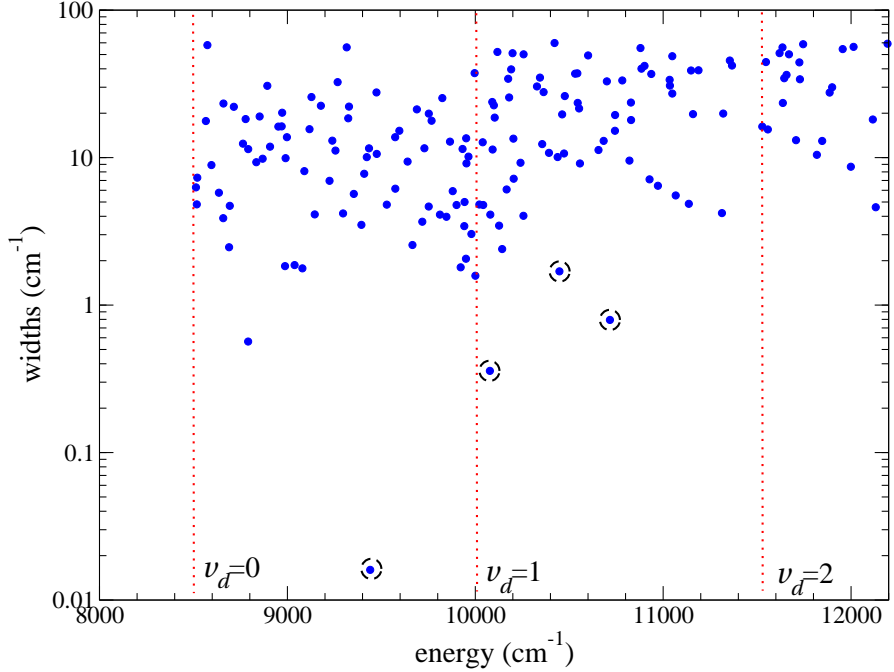}
\caption{Widths, $\Gamma$, of  resonances of the $A_2 (D_{3h})$ vibrational levels. Vertical dotted lines indicate threshold energies for dissociation channels with different excitation of the oxygen molecule $v_d = 0, 1,$ and 2. Numerically, lifetimes $\tau$ in ps are related to the widths in \cm  as $\tau[\mathrm{ps}]=(2\pi c \Gamma[\mathrm{cm^{-1}}])^{-1}$, where $c$ is the speed of light in units of cm/ps, $c=0.0299792458$ cm/ps. Wave functions of the encircled levels are shown in Figs.~\ref{fig:Jacobi_A2_v1_diss} and~\ref{fig:Jacobi_A2_diss}. }
\label{fig:resonances_A2}
\end{figure}

\begin{figure}[ht]
\vspace{1cm}
\includegraphics[width=12cm]{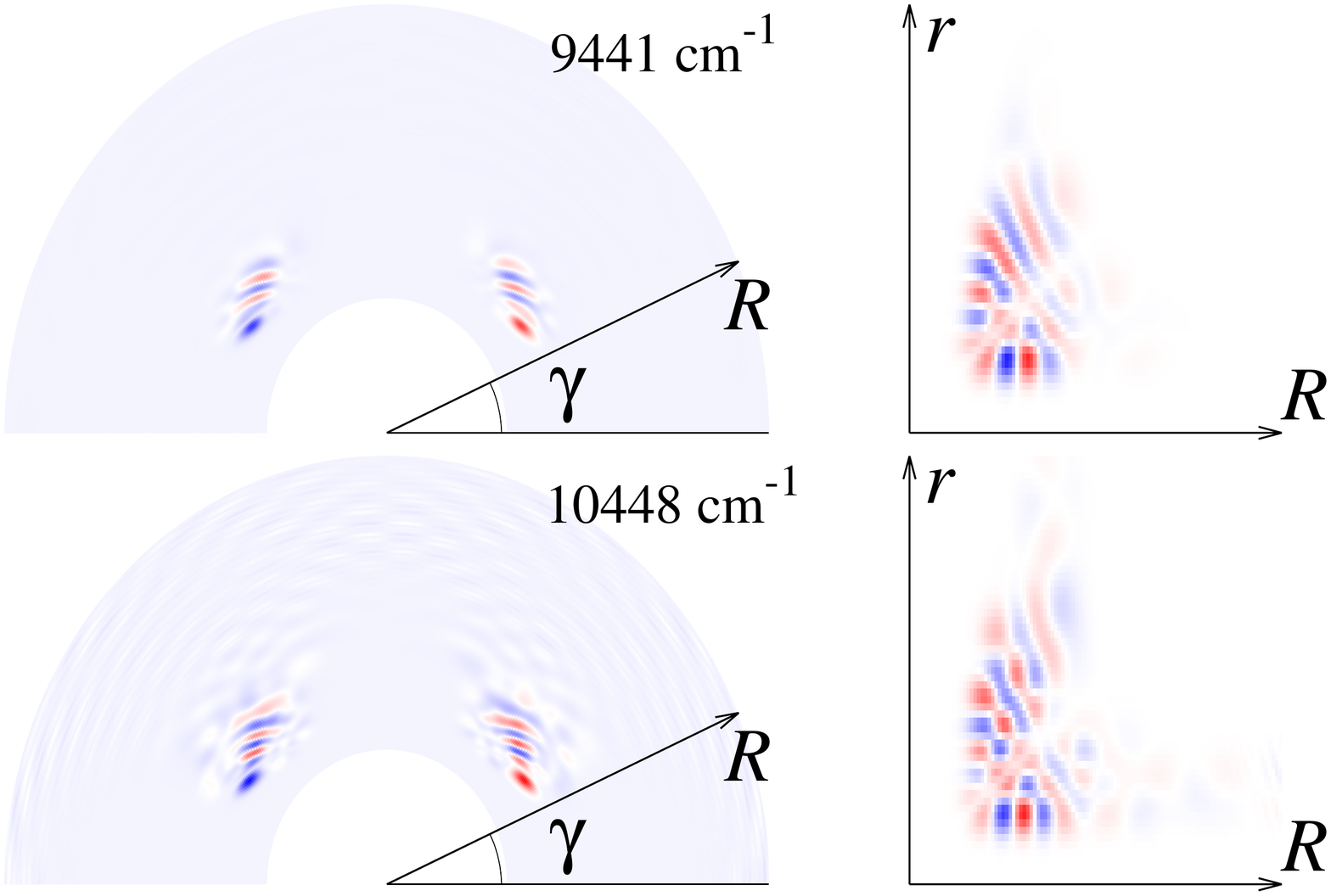}
\caption{Vibrational part of the wave functions of the $(8,0,1)$ (upper panels) and $(9,0,1)$ (lower panels) levels of $A_2(D_{3h})$  vibrational symmetry. The calculated widths are $\Gamma=0.016$~\cm for the $(8,0,1)$ level and and 1.7~\cm for $(9,0,1)$. }
\label{fig:Jacobi_A2_v1_diss}
\end{figure}

\begin{figure}[ht]
\vspace{1cm}
\includegraphics[width=12cm]{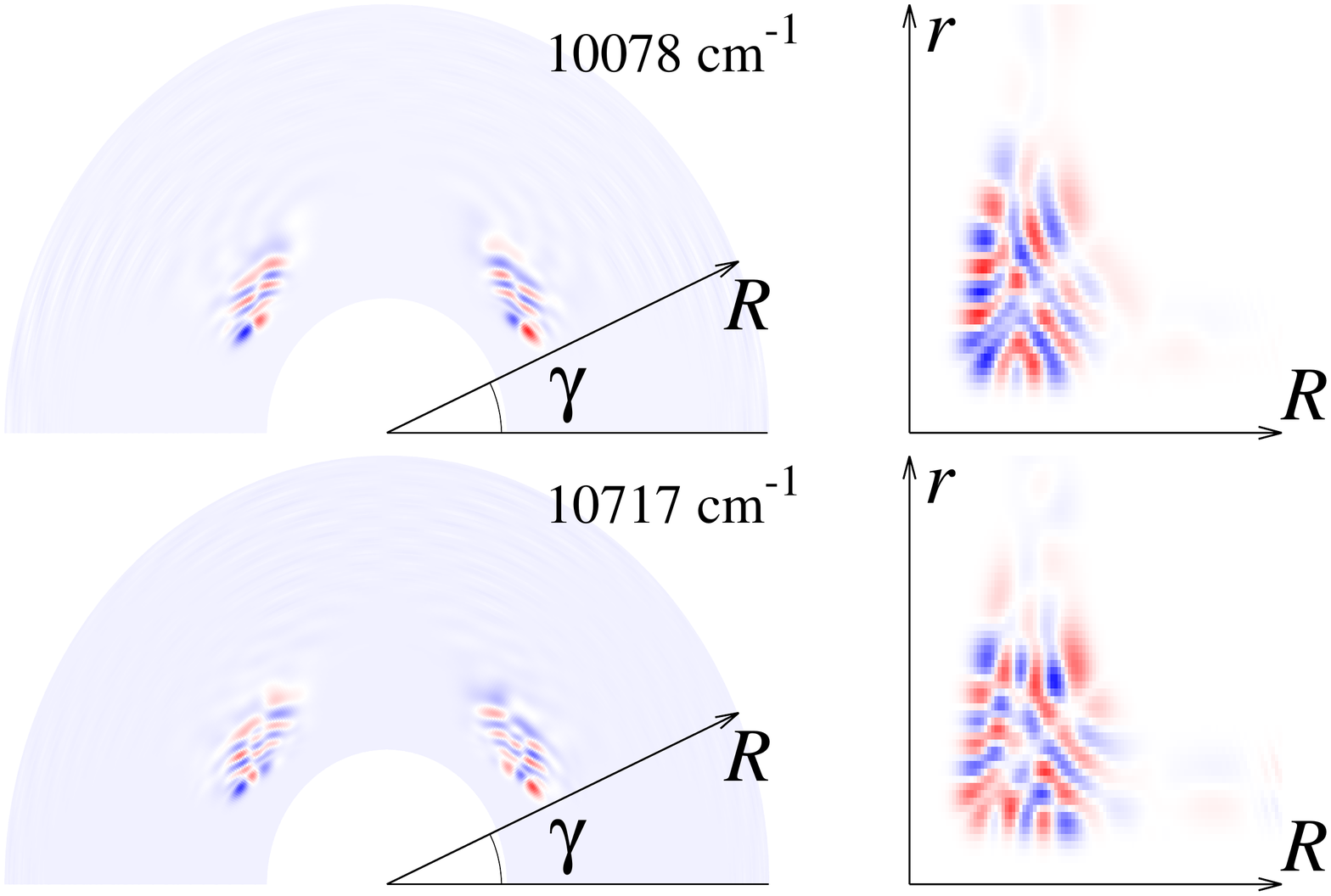}
\caption{Vibrational part of the wave functions of two highly-excited levels of the $A_2(D_{3h})$ symmetries. The calculated widths are $\Gamma=0.36$~\cm for the function shown in the upper panels and 0.8 ~\cm for the function shown in the lower panels. }
\label{fig:Jacobi_A2_diss}
\end{figure}

Figure~\ref{fig:Bv_A2} shows the vibrational dependence of the rotational constants $A_v$ obtained for $A_2$ vibrational symmetry in the $D_{3h}(M)$ group. The energy origin of the figure is the same as in  Fig.~\ref{fig:Bv_A1}, i.e. the energy of the ground rovibrational level of ozone $(0,0,0),J=0$. The same three families of vibrational levels corresponding to the three normal modes, are easily identified. The figure also includes some of the low-energy predissociated resonances above the dissociation limit. Figure \ref{fig:resonances_A2} shows widths of the $A_2$ vibrational levels situated above the dissociation threshold. Most of the resonances shown in the figure have widths between 2 and 70~\cm (lifetimes between 0.08 and 2~ps) with a few outliers having significantly smaller widths. These outliers are the levels highly-excited in the $v_1$ mode, as demonstrated in Figs.~\ref{fig:Jacobi_A2_v1_diss} and Figs.~\ref{fig:Jacobi_A2_diss}.

\begin{figure}[ht]
\vspace{1cm}
\includegraphics[width=12cm]{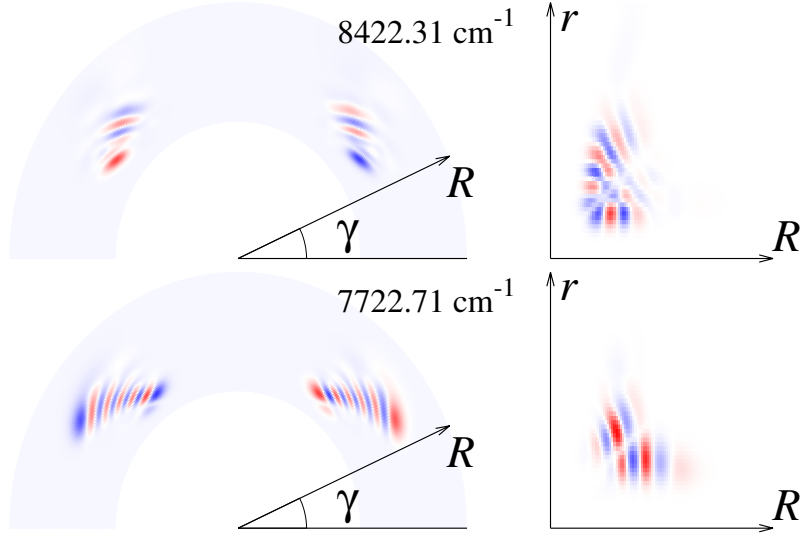}
\caption{Same as Fig.~\ref{fig:Jacobi_A1_v1_v2}, but for the $(7,0,1)$ and $(0,10,1)$ wave functions, of $A_2(D_{3h})$ overall vibrational symmetry.}
\label{fig:Jacobi_A2_v1_v2}
\end{figure}

Figures~\ref{fig:Jacobi_A2_v1_v2}, \ref{fig:Jacobi_A2_v1_diss}, \ref{fig:Jacobi_A2_diss} shows some of the bound and resonance vibrational levels of $A_2$ vibrational symmetry of the $D_{3h}$ group. The vibrational levels $(v_1,v_2,v_3)$ with odd $v_3$ have overall $A_2$ symmetry. As mentioned above, continuum states (including dissociative states) of ozone $^{16}$O$_3$ can only be of $A_2$ vibrational symmetry. 
Figure~\ref{fig:Jacobi_A2_v1_diss} demonstrates two resonance wave functions from the $(v_1,0,1)$ series. Although excitation of the $v_1$ mode differs for these two levels only by one quantum, their lifetimes are very different, 330~ps for the  $(8,0,1)$ and 3.1~ps for the $(9,0,1)$ level. Figure~\ref{fig:Jacobi_A2_diss} shows two examples of wave functions for levels where all three modes are excited and mixed.

\begin{figure}[h]
\vspace{1cm}
\includegraphics[width=12cm]{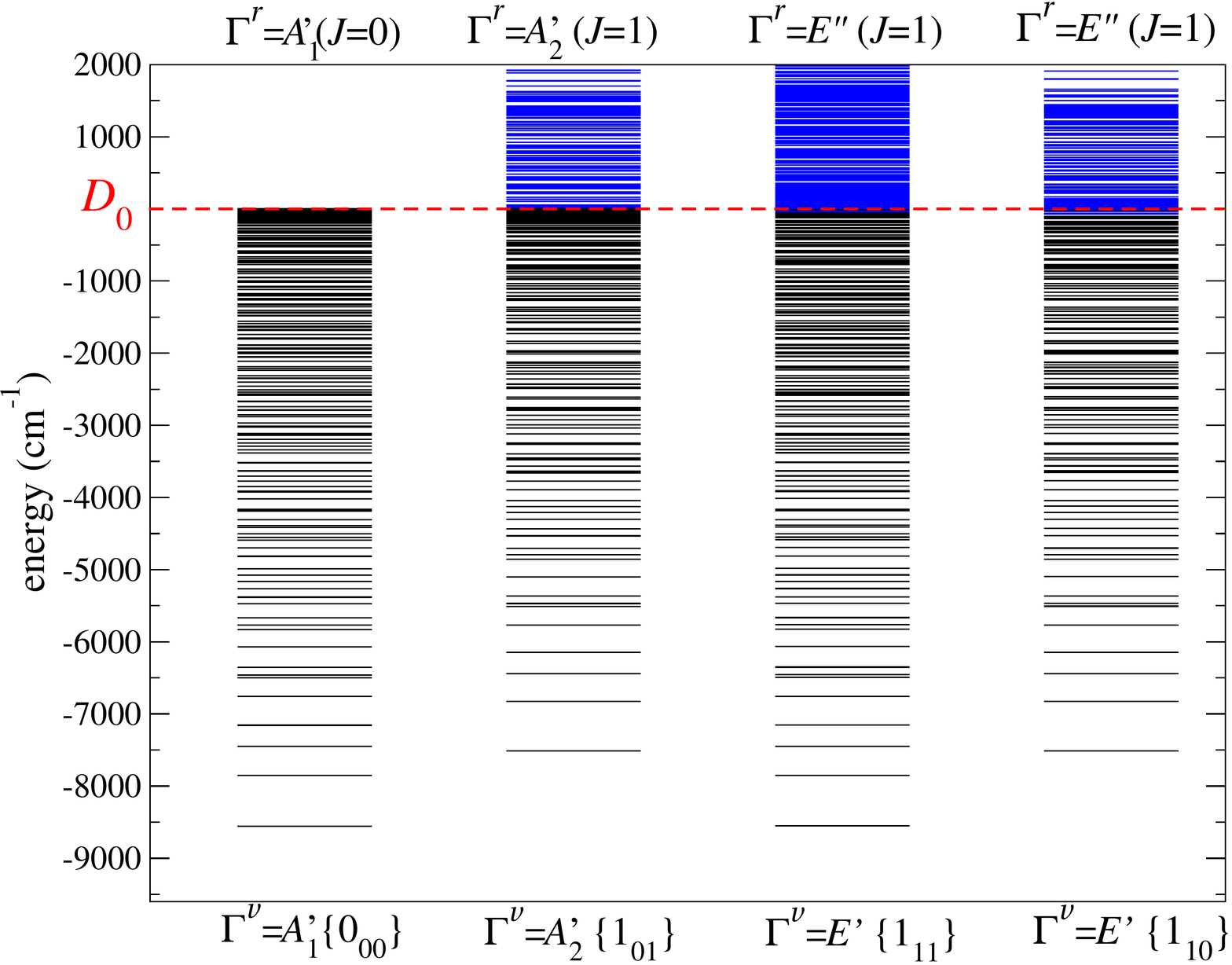}
\caption{Energies of vibrational levels for $J=0$ and $J=1$. In addition to the symbol $\{J_{K_aK_c}\}$ of the rotational states of an asymmetric top molecule, irreducible representations of the vibrational part ($\Gamma^v$, below the graph) 
and rotation part ($\Gamma^r$, above the graph) of the wave function are  specified. 
The corresponding resonances are shown in blue color.}
\label{fig:energy_levels}
\end{figure}

Figure \ref{fig:energy_levels} shows the energies of symmetry-allowed levels for the two lowest values of the angular momentum, $J=0$ and 1. The standard notation notation $\{J_{K_aK_c}\}$ for rotational states of an asymmetric top molecule is used for bound states within the well. The irreducible representation in the $S_3$ permutation group of vibrational part of the total wave function is also specified. Only states of  $A_2 (D_{3h})$ vibrational irrep can dissociate and, therefore, only resonances of this symmetry are shown in the figure.

\section{Conclusion}

In this study, energies, widths, and wave functions of $^{16}$O$_3$ vibrational resonances   were determined for levels up to about 3000 \cm above the dissociation threshold. The predissociated resonances have lifetimes between 0.08 and 2~ps with a few long-living levels.  These outliers are levels with the highly-excited $v_1$ and $v_2$ modes. An example of a long living state is (9,0,0) $J$=1 level with the lifetime of 330~ps.  Energies of bound states of the ozone isotopologue $^{16}$O$_3$ up to the dissociation threshold were also computed. The total permutation inversion symmetry $S_3\times I$ of the three oxygen atoms was taken into account using hyper-spherical coordinates. The effect of the symmetry is negligible for the levels deep in the ozone potential, but vibrational levels near the dissociation  threshold cannot be represented correctly within one potential well and, therefore, the complete permutation symmetry group should be used. 

Symmetry properties of allowed rovibrational levels of ozone (applicable to $^{16}$O$_3$ and $^{18}$O$_3$) as well as correlation diagrams between the bound-state and dissociation regions were derived and discussed. The correlation diagrams are not trivial because ozone dissociates to (or is formed from)
a $P$-state oxygen atom and an O$_2$ molecule of symmetry $^3\Sigma^-_g$. 

Within the employed model including only the lowest PES of ozone, the purely vibrational states, i.e. $J=0$ states, of ozone $^{16}$O$_3$  (and  $^{18}$O$_3$)  
cannot dissociate  to the fragments allowed by symmetry of the electronic ground state of the O$_2$ molecule. Note that excited rotational states with $J>0$ satisfying Eq.~(\ref{eq:scattering}) do exist. Examples of such resonances are shown in Figs.~\ref{fig:Jacobi_A2_v1_diss} and \ref{fig:Jacobi_A2_diss}.  We would like to stress here, that the single electronic PES model neglects the coupling of the angular momentum of the molecular frame, $\boldsymbol R$, with the electronic angular momentum, $\boldsymbol L$, which is not zero. 
In general, the total (but without nuclear spin, $\boldsymbol I$) angular momentum $\boldsymbol J$ can be written as 
$\boldsymbol J = \boldsymbol R + \boldsymbol L + \boldsymbol S + \boldsymbol \Pi$, where $\boldsymbol S$ is the electronic spin and $\boldsymbol \Pi$ 
the vibrational angular momentum. From this we obtain the approximate quantum number of the rotation of the molecular frame as $\boldsymbol R = \boldsymbol J - \boldsymbol L - \boldsymbol S - \boldsymbol \Pi$. 
Neglecting the effect of $\boldsymbol L$ and $\boldsymbol S$ in the rovibrational problem, 
$\boldsymbol R \approx \boldsymbol J$ is a ``good'' quantum number. Our rovibrational energies have been calculated within this approximation, as have been those obtained by other workers in the field.
However, the importance of the electronic angular momentum is evident from asymptotic behavior of Eq.~(\ref{eq:scattering}): 
At large distances between O and $\rm O_2$, the electronic angular momentum is clearly not zero. In a more accurate model, the electronic momentum should be
accounted for and coupled to the angular momentum of the nuclear frame, due to the cross-terms generated by $\boldsymbol R^2$, conserving the total angular momentum, $\boldsymbol J$. 
In such a more accurate model, 
the continuum vibrational spectrum for $J \approx R = 0$ is allowed (since $R$ is not ``conserved''  any more).  The corresponding vibrational resonances should have relatively long lifetimes because they can only decay due to non-Born-Oppenheimer and Coriolis couplings involving the three PES's converging to the same dissociation limit, with the oxygen atom being in the triply-degenerate electronic state.  Such long-living states above the dissociation threshold, for example, $(9,0,0)$ and $(10,0,0)$ have indeed been observed in experiments. Therefore, an accurate theoretical determination of lifetimes of $J=0$ resonance levels would involve three potential energies surfaces.  

The above discussion did not take into account spin-orbit coupling. For even more realistic description of the nuclear motion states in the continuum, one has to consider
the effect of coupling of the electronic singlet state with electronic triplet states, of symmetry $B_1$ and $A_2$ in $C_{2v}(M)$, or $A_2''$ and $A_1''$ in $D_{3h}(M)$,
or $A''$ in $C_s(M)$, that approach the same asymptotic dissociation limit, O$(^3P)$ 
and $\rm O_2$($X^3\Sigma_g^-$), as the electronic ground state.  
Rosmus, Palmieri and Schinke~\cite{ROS02:4871} have determined the spin-orbit coupling elements with 
all relevant triplet states in the asymptotic channel.
The matrix elements are of the order of $\langle X\, ^1A_1' | H_{so} | ^3A_2'' \rangle \approx 60 \, \cm$. The long-range behavior of the potential energy surfaces accounting for spin-orbit coupling was discussed in Ref.~\cite{lepers2012long}.

To study the effect of spin-orbit coupling, the total nuclear-electronic wave function 
must be expanded, including for simplicity just one generic triplet state, as
\begin{equation}
\label{eq:so-expansion}
\Psi_{vJm}(\Omega,{\cal Q})=\sum_k \left[ c_1 \, |^1A_1'\rangle  \psi_{vJk; ^1A_1'}({\cal Q})
+ c_2 \, |^3A_2''\rangle  \psi_{vJk; ^3A_2''}({\cal Q}) \right] {\cal R}_{Jkm}(\Omega)\,.
\end{equation}
As before, the product of electronic and nuclear motion functions must have the same symmetry 
as the rotational function, except for their parity. However, the nuclear motion component of 
the $^3A_2''$ electronic state is antisymmetric and can therefore correlate with the 
asymptotic $\rm O_2 \, + \, O$ wave function. A full treatment of the nuclear dynamics of ozone accounting for the spin-orbit coupling would involve solving the rovibrational Schr\"odinger equation on several coupled potential energy surfaces, which is hardly possible at present. However, an adiabatic approach with respect to the spin-orbit coupling should also be accurate and could be used in a future study. In the approach, the first step would be to construct the matrix of the potential energy. The matrix would include the lowest three Born-Oppenheimer PES's, mentioned above, and the spin-orbit coupling such as described in Refs.~\cite{ROS02:4871,lepers2012long}. The matrix then should be diagonalized for each geometry, which will produce adiabatic potential surfaces accounting for the spin-orbit coupling. Because the lowest Born-Oppenheimer PES ($X^1A_1$) does not cross the two other PES's at energies near or below the dissociation threshold, after the diagonalization, the lowest obtained PES will be very similar to the original $X^1A_1$ Born-Oppenheimer PES, except that the dissociation limit will be shifted down. Low-energy rovibrational states  obtained with the new adiabatic PES will be almost identical to the ones discussed in this study. States near the dissociation threshold (approximately 60~\cm above and below $D_0$) will have energies somewhat different compared to the ones obtained with the PES without the spin-orbit coupling. However, qualitatively the structure of rovibrational levels near the dissociation will stay the same.

\section*{Acknowledgments}
This work is supported by the CNRS through an invited professor position for V. K. at the GSMA,
the National Science Foundation, Grant No PHY-15-06391
and the ROMEO HPC Center at the University of Reims Champagne-Ardenne. The supports from Tomsk State University Academic D. Mendeleev
funding Program, from French LEFE Chat program of the Institut National des Sciences de l'Univers (CNRS) and from Laboratoire International Franco-Russe SAMIA are acknowledged.

\clearpage

%

\end{document}